\documentclass [12pt]{article}
\RequirePackage[english]{babel}
\RequirePackage[cp1251]{inputenc}
\usepackage{amsmath,amssymb}

\begin{document}
\sloppy
\begin{center}
{\bf \Large{Enhancement Mechanisms  of Low Energy Nuclear Reactions
}\\
\vspace*{0.5cm}
\large{Gareev F.A.,  Zhidkova I.E.}}\\
\vspace*{0.2cm}
{\sl Joint Institute for Nuclear Research,\\ 141980, Dubna, Russia}\\
gareev@thsun1.jinr.ru\\
zhidkova@jinr.ru
\end{center}
\section{Introduction}

 One of the fundamental presentations  of nuclear physics  since the very early
 days of its study has been the common assumption  that the
 radioactive process (the half-life or decay constant) is
 independent of external conditions. Rutherford, Chadwick and
 Ellis \cite{RUT30} came to the conclusion that:
\begin{itemize}
\item «the value of $\lambda$ (the decay constant) for any substance is a characteristic constant independent of all
physical and chemical conditions».
\end{itemize}

This very important conclusion (still playing a negative role in
cold fusion phenomenon) is based on the common expectation (P.
Curie suggested that the decay constant is the etalon of time) and
observation that the radioactivity is a nuclear phenomenon since
all our actions affect only states of the atom but do not change
the nucleus states. We cannot hope to mention even a small part of
the work done to establish the constancy of nuclear decay rates.
For example, Emery G.T. stated \cite{EME72}:
\begin{itemize}
\item «Early workers tried to change the decay constants of various
members of the natural radioactive series by varying  the
temperature between $24^{\circ}K$ and $1280^{\circ}K$, by applying
pressure of up to 2000 atm, by taking sources down into mines and
up to the Jungfraujoch, by applying magnetic fields of up to
83,000 Gauss, by whirling sources in centrifuges, and by many
other ingenious techniques. Occasional positive results were
usually understood, in time, as result of changes in the counting
geometry, or of the loss of volatile members of the natural decay
chains. This work was reviewed by Meyer and Schweider
\cite{MEY27}, Kohlrausch \cite{KOH28}, and Bothe \cite{BOT33}.
Especially interesting for its precision is the experiment of
Curie and Kamerlingh Onnes \cite{CUR13}, who reported that
lowering the temperature of radium preparation to the boiling
point of liquid hydrogen changed its activity, and thus its decay
constant, by less than about $0.05\%$. Especially of Rutherford
and Petavel \cite{RUT07}, who put a sample of radium emanation
inside a steel-encased cordite bomb. Even though temperatures of
$2500^{\circ}C$ and pressures of 1000 atm were estimated to have
occurred during the explosion, no discontinuity in the activity of
the sample was observed».
\end{itemize}

It seems (in that time) that this conclusion was supported by the
following very simple and strong
arguments (Common Sense):\\
1. Nuclear processes have characteristic energies $\approx 1$ MeV,
whereas chemistry has a few eV per atom, molecules have a part of
eV. The inner atomic shells are bound with many keV in the medium
and heavy
elements.\\
2.The localization of electrons in atoms is $\approx 10^{-8}$ cm,
whereas the localization of nucleons in nuclei is $\approx
10^{-13}$ cm.\\
 Therefore, the nucleus should be unaffected by superficial atomic
changes: nuclear processes should not be influenced by the
surroundings. The constancy of nuclear decay rates was firmly
established, confirming evidences from experimental studies of
$\alpha$- and $\beta$-decays and theoretical estimations.

The constancy of nuclear decay rates acquired the strength as a
classical law.  Any papers contradicting this law were ignored by
all the scientific journals as erroneous ones.

The history of science has own laws. The ground of the
$\beta$-decay of nuclei was given by E. Fermi in 1934 year. It was
very easy to prove that certain processes of radioactive decay
should be intimately connected with the presence of atomic
electrons and may be affected by the changes in the electronic
structure produced by chemical compounds. It took 13 years to
understand this very simple phenomenon. The possibility of
altering the decay rate of $Be^{7}$ was suggested in 1947 by Segre
\cite{SEG47} and by Dodel \cite{DOD47,DOD47A}. In the case of
electron- capture decays the decay rate is directly related to the
density of atomic electrons in the nucleus and the effects of
different chemical environments should be measurable. The
theoretical foundation was the following \cite{SEG47}:\\
\begin{itemize}
\item «The radioactive decay constant of  a substance decaying by
orbital electron capture is proportional to $\mid\psi(0)\mid^{2}$
of the electrons. In the case of a light element like $^{7}Be$ it
may be possible to alter this quantity by an appreciable amount by
putting the $Be$ in different chemical compounds. We would then
have a slight change of the radioactive half-life of the $Be$ in
different compounds. The magnitude of the effect may be in the
neighborhood of one percent, but it is practically impossible to
give quantitative estimate because the total change of $\psi(0)$
is affected  by certain factors such as the density of the
crystal, nature of the chemical bond, etc. they are both positive
and negative, and have comparable magnitudes. To obtain a reliable
estimate of the effect we require a more detailed knowledge of the
wave functions for various compounds than is at present available.
Experiments are in progress to detect the effect by comparing the
half-life of $^{7}Be$ in $Be$ metal with that in $BeO$ or
$BeF_{2}$».
\end{itemize}

The confirmed altering decay rate for $^{7}Be$ in different
chemical compounds  was of the order $0.1\%$ \cite{EME72}. The
6-hr isomer $^{99m}Tc$ decays by internal conversion of a 2.2-keV
E3 transition. The observed effects in different chemical forms
were of the order $0.3\%$ \cite{EME72}. The greatest chemically
induced half-life changes of the order $3.5\%$ were established in
\cite{COO65}.

The half-life of $^{7}Be$ electron capture was measured
\cite{OHT04} in endohedral fulleren $^{7}Be@C_{60}$ and $^{7}Be$
metal: $T_{1/2}=52.68\pm 0.05$ and $T_{1/2}=53.12\pm 0.05$ days,
respectively. This $0.83\%$ difference between the electron
capture in $C_{60}$ and in $^{7}Be$ metal represents a strong
environmental effect on the $^{7}Be$ EC capture rate, caused by
the different electronic wave functions near the $^{7}Be$ nucleus
inside a $C_{60}$ cage and inside $Be$ metal.

A weak interaction which is responsible for electron capture and
other forms of beta decay is of very short range. So the rate of
electron capture and emission (internal conversion) is
proportional to the density of electrons at the nucleus. It means
that we can manage the electron-capture (emission) rate by the
change of the total density in the nucleus. It can be carried out
in different macroscopic  ways by using available environmental
effects. These questions were highlighted in different reviews and
books \cite{EME72,SER66,STA69, KON66,SCHO66,WU66,BOU60,BER68} at
the end of the seventies of the 20th century. The reader should
compare the common by accepted conclusions about the decay rates
in the thirties and seventies of the 20th century.

Data on pleochroic halos led to the conclusion \cite{SPE72} that
these data do not provide a convincing proof that the laws of
radioactive decay are constant in time.

Shnol S. and coauthors  \cite{SHN98} came to the conclusion that
the decay rates of radioactive nuclei change in time with the
period of 24 hours, 27, and 365 days. Periodic variations in
$\beta$-decay rates of $^{60}Co$, $^{90}Sr$ and $^{137}Cs$ were
discovered \cite{BAU95,BAU96,BAU98,BAU01}. The 27-day and 24-hour
period in these changes were found.

The aim of this talk is to discuss the possibility of inducing and
controlling nuclear reactions at low temperatures and pressures by
using different low-energy external fields and various physical
and chemical processes. The main question is the following: is it
possible to enhance LENR rates by using the low and extremely low
energy external fields?

\section{Cold fusion and transmutation}

In 1989 Fleishmann M. and Pons S.  reported about their
observation of nuclear products and excess heat on a palladium
electrode during the electrolysis of solutions in heavy water. The
electrochemical experiments were interpreted by the authors as a
result of nuclear fusion reaction (named cold fusion) but the
scientific community rejected this interpretation. More than 3000
papers in the field of cold fusion and transmutation (further the
low-energy nuclear reaction LENR) were published. Various
anomalous results were observed at low temperatures and pressures
which are beyond the framework of modern theoretical paradigm. The
theoretical models are not able describe these anomalies even
qualitatively. The reader can find the history and problems of
cold fusion in the Proceedings of the International Conferences on
Condensed Matter Nuclear Science, the Russian Conferences on Cold
Nuclear Transmutation of Chemical Elements and Ball Lighting, and
also in a recent review of the Department of Energy of the USA
\cite{HAG04} and books \cite{KRI04, KIR04}. Russian experimental
data on the low-energy nuclear reactions are published in
\cite{URU00,URU02,BAL03,KUZ03} and their new theoretical
interpretation was given in \cite{GAR04,GAR04A,GAR03,GAR02}.

The general important conclusion can be drawn from the studies
performed  during 15 years:
\begin{itemize}
\item The poor reproducibility of experimental results  and
extreme difficulties of their interpretation in the  framework of
modern standard theoretical physics are the main reasons of the
persistent nonrecognition of cold fusion and transmutation
phenomenon.
\end{itemize}

Recent progress in both directions is remarkable (see Abstracts
ICCF-11, Marseille: France: 2004, 31 October - 5 November); in
spite of being rejected by physical society, this phenomenon is a
key point for further success corresponding fundamental research.

\subsection{Reproducibility of Low-Energy Nuclear Reaction
Experiments}

Reproducibility of experiments within and between laboratories is
a fundamental requirement and cornerstone for any scientific
investigations. There are many fundamental factors  that are
relevant to the issue of reproducibility (for details see
\cite{NAG03}).

Everybody with a perfect ear will say that, for example, on the
viola play will have never be reproducible: it depends on too many
factors (resonance conditions) which are impossible to repeat. The
semiconductor effect in a transistor is extremely sensitive to
damages and impurities of crystal which were impossible to control
in the initial experiments. The degree of reproducibility was
increasing over the years when the properties of used materials
were improved and standardized, and the process was optimized and
controlled with high accuracy. The same would happen for
reproducibility of LENR \cite{KRI04}. We will show that
expectation for LENR is correct only partly\footnote{References to
original and review cold fusion literature are not given in our
talk. They are available in the Proceedings of ICCF.}.

The targets in  standard nuclear physics using accelerators are
the substances in the ground states: the gases, amorphous solids
or crystalline solids. The projectile particle interaction with
target nuclei has taken place in vacuum. Therefore, the influence
of the surrounding matter (say, atomic electrons) on the velocity
of such nuclear processes (especially at high energies) should be
negligible. It seems that these expectations supported by
estimations of energy and size differences ($10^{-5} - 10^{-6}$)
of atoms and nuclei and experiments show almost a full
reproducibility.

We come to the following conclusion:
\begin{itemize}
\item A greater part of processes in nuclear physics takes place
in  closed systems. Reproducibility of such experiments should be
independent of the place and time of measurements - a cornerstone
of the modern scientific method.
\end{itemize}

LENRs occur in the surroundings (gases, condensed mater, water,
solutions,..) which are induced by low-energy  external fields as
ultrasounds, electromagnetic fields, lasers,... So atoms,
molecules in the surroundings and  atoms of interacting nuclei are
in excited states or ionized. Nuclei, atoms, the surrounding
medium, and external fields representing  interacting subsystems
are form a dynamical open system. Frequencies and phases of
subsystem motions may be coordinated according to the universal
resonance synchronization principle (see Appendix) and the result
may be a creation of a collective (coherence) state for the whole
system. We will call such a system an auto-oscillation system in
which the frequencies of an external field and frequencies of the
all subsystems are commensurable. The demand for frequency
commensurability means that all motions in a system are in
co-ordination (in resonance), which is difficult to fulfil. This
is the cause of poor reproducibility of LENR.

We formulate as a working hypothesis the following assumption:
\begin{itemize}
\item LENRs take places in open systems in which all frequencies and
phases coordinated according to the universal resonance
synchronization principle -- the main reason for poor
reproducibility. Poor reproducibility and unexplained results do
not mean that the experiment is wrong.
\end{itemize}

\subsection{The Bound State $\beta^{-}$ Decay}

Bound state $\beta^{-}$ decay ($\beta_{b}$), in which the decay
electron remains in an electron bound state of the daughter atom
and the monochromatic antineutrino carries the total decay energy
$Q$, was first predicted by  R. Daudel, M. Jean, and M. Lecoin
\cite{DAU47} in 1947 and discussed in
\cite{BAH61,BAT76,KOP86,TAK83,MAT87}.

This new decay mode, the bound state $\beta^{-}$ decay, was for
the first time experimentally observed for bare $^{163}Dy$
\cite{JUN92} (Bare means that the atom $^{163}Dy$ is ionized
fully. We will use the designation for this case as
$^{163}Dy^{66+}$) and $^{187}Re$ \cite{BOS96}. Nucleus $^{163}Dy$
is stable as a neutral atom ($Q_{\beta}=-2.565$ keV) and become
radioactive when fully ionized atoms (bare nucleus
$^{163}Dy^{66+}$) decay to $^{163}Ho^{66+}$
($Q^{K}_{\beta_{b}}$=+50.3 keV)   via the bound state $\beta_{b}$
decay with a half-life of 47 days. Nucleus  $^{163}Ho$ is unstable
and it is transferred to $^{163}Dy$ by electron capture with a
half-life of $4.6 \cdot 10^3$ yr. Difference of masses
$m(^{163}Ho) - m(^{163}Dy) =2.6$keV; therefore, the electron
capture is only possible from M- or higher orbits.Unstable nuclei
$^{163}Ho$  become practically stable due to ionization of atoms
$^{163}Ho$  up to these orbits because the electron capture in
these cases is only possible from continuum states which have an
extremely small probability. The ionization of atoms changes the
beta decay direction of nuclei: in neutral atoms $^{163}Ho$
($^{163}Dy$) the electron capture leads to the transition
$^{163}Ho \rightarrow ^{163}Dy$($^{163}Dy$ are stable), in fully
ionized atoms bare nuclei $^{163}Ho^{67+}$ ($^{163}Dy^{66+}$)
are stable (unstable).
\begin{itemize}

\item      General conclusion: in neutral atoms  some ground state
nuclei  decay via orbital electron
      capture, for bare nuclei (fully ionized atoms) the electron capture branches are blocked.
      In these cases (if in addition the positron decays are lacking) bare nuclei become stable.
      This conclusion is  very strong and wells-known in nuclear society.
\end{itemize}

For neutral $^{187}Re^{0+}$ only a unique, first forbidden
transition to the $^{187}Os$ ground state is energetically
possible. The small matrix element and the small $Q_{\beta}$ value
of $Q_{\beta}$=2.663(19) keV lead to the long half-life of 42 Gyr.
The measured half-life \cite{BOS96} for bare $^{187}Re^{75+}$
($Q^{K}_{\beta b}$=+72.97 keV)  of $T_{1/2}=(32.9\pm 2.0)$ yr is
billion times shorter than that for neutral $^{187}Re$.

\begin{itemize}
\item The ground state $\beta$ decay properties of nuclei
cardinally change when all electrons of the atomic shells are
removed: stable (unstable) nuclei become unstable (stable) and a
half-life may decrease up to billion times - 9 orders of
magnitude. The interpretation is very simple:  magnitude of
$Q_{\beta}$  and phase volume increase for the ionized atoms
rather than for neutral ones.
\end{itemize}

\subsection{Nuclear Decay of Coulomb Excited and Isomeric States for Fully
Ionized Atoms}

The half-lives of isomeric states of fully ionized
$^{144m}Tb^{65+}$, $^{149m}Dy^{66+}$ and $^{151m}Er^{68}$ were
measured \cite{LIT03}. The increase was observed of the half-lives
of bare isomers by factors of up to 30 to their neutral
counterparts. The authors  \cite{LIT03} give the correct and
evident interpretation of experimental results:

\begin{itemize}
\item This is due to the exclusion of the strong internal
conversion and electron-capture channels in the radioactive decay
of these bare nuclei.
\end{itemize}

Experiments with highly-ionized $^{57}Fe^{q+}(q=19-25)$
projectiles at 6 MeV \cite{PHI93} and $^{125}Te^{q+}(q=46-48)$
projectiles at 27 MeV/u \cite{ATT97} have demonstrated a growth
(ranging from a few $10\%$ up to $670\%$) of nuclear half-lives of
Coulomb excited levels due to the direct influence of the
electronic configuration on the internal-conversion coefficients.

\subsection{The Effect of Host on the Half-life
of $^{7}Be$}

Norman E.B. et al. \cite{NOR01} measured $^{7}Be$ decay rates in
gold ($Au$), graphite, boron nitride and tantalum ($Ta$). Among
these materials , they found that the $^{7}Be$ half-life was the
longest in $Au$ and the shortest in graphite. According to their
experiments, the decay rate of $^{7}Be$ in $Au$ is lower than that
in graphite by $(0.38\pm 0.09)\%$.

Ray A. \cite{RAY99} measured the difference of $^{7}Be$ decay
rates in $Au$ and $Al_{2}O_{3}$ and found that the decay rate in
$Au$ was lower than that in $Al_{2}O_{3}$ by $(0.72\pm 0.07)\%$.

Ray A. et al. \cite{RAY02} pointed out that the apparent
disagreement between the two sets of experimental results was most
likely due to the choice of different reference samples with which
the comparisons were carried out. Indeed, Norman E.B. et al.
\cite{NOR01} used the $^{7}Li$ beam for their implantation
studies, whereas Ray A \cite{RAY99} used the proton beam. The
radiation damage by $^{7}Li$ on gold lattice sites, where $^{7}Be$
nuclei stop, would be much larger \cite{ZIE85} ($3*10^{-4}$
vacancies/Angstrom/ion) than the corresponding damage ($10^{-5}$
vacancies/angstrom/ion) for proton. Therefore, the radiation
damage effects on lattice due to heavy ion irradiation might also
be partly responsible for apparent discrepancies. It means that to
speak about reproducibility in this case we should take into
account at least atomic physics effects that are usually ignored.

The ratio of $L$ to $K$-shell electron capture in $^{7}Be$ bare
nucleus shows \cite{VOY02} that the measured ratio is less than
half of the existing data for free $^{7}Be$.

\begin{itemize}
\item These discrepancies are most likely due to the distortions of $L$
and $K$-shell orbitals by the host medium.
\end{itemize}

\subsection{Controlled Gamma-Decay of Excited Nuclei}

According to the modern theory, the spontaneous gamma-decay of
excited nuclei in free space without any material bodies is a
noncontrolled process. Probability $A_{eg}$ of this decay

$$A_{eg}\equiv
\frac{1}{\tau}=\frac{4\pi^{2}\omega_{eg}\mid\vec{d}_{eg}\mid^{2}\rho(\omega_{eg})}{3\hbar}=
\frac{4\omega_{eg}^{3}\mid\vec{d}_{eg}\mid^{2}}{3\hbar c^3}$$ is
fully determined by the matrix element $\vec{d}_{eg}$ of the
nucleus dipole moment  and  energy of the nuclear transition
$h\omega_{eg}=E_{e}-E_{g}$.

The total lifetime $\tau_{tot}=\tau/(1+\alpha)$ and radiative
lifetime $\tau$ of this excited nucleus in free space are the
constants. Here $\alpha$ is the coefficient of internal electron
conversion for the nuclear transition $E_{e}-E_{g}$.

Problems become very complicated in the important case  when
material bodies are present in the surrounding space. Vysotskii
V.I. \cite{VYS98} considered the general system which included the
excited atom nucleus, the system of atom electrons, the system of
zero-energy (in vacuum state) electromagnetic modes, and the
screen -- the system of $N$ resonant or nonresonant atoms situated
at the distance $d\gg\lambda_{eg}=2\pi c/\omega_{eg}$. The authors
of \cite{VYS01} concluded that

\begin{itemize}
\item It is usually stated that in all cases with the presence of any
material bodies at a macroscopic distance $d\gg\lambda_{eg}=2\pi
c/\omega_{eg}$ from the excited nucleus the expression for the
lifetime $\tau$ and $\tau_{tot}$ remains the same or changes by an
unmeasurable small value. Such a supposition is erroneous. It was
shown \cite{VYS98} that a spontaneous gamma-decay was a process of
an excited nucleus relaxation, the phase promise of which was
caused by interaction with a fluctuating state of the thermostat
at the distance $d\gg\lambda_{eg}$ from the nucleus. The
phenomenon of a controlled nucleus gamma-decay is a result of
interaction of the nucleus with zero-energy modes, interaction of
these modes with the atoms of controlled (and controlling) screen,
and interaction of the nucleus with the system of atom electron.
\end{itemize}

The increase in radiative lifetime $\tau$ of an excited nucleus by
$10-40\%$ and total lifetime $\tau_{tot}$ by $1\%$ was observed in
the experiments \cite{VYS01,VYS97,VYS98A} with gamma-source
$^{57}Co(^{57*}Fe)$ and with gamma-absorber made of stable
$^{57}Fe$ isotope. So these results prove the possibility of
controlled essential influence of a thin resonant screen on the
amplitude, space and temporal characteristics of a spontaneous
decay and excited nuclei radiation.

\subsection{Okorokov Effect}

Let us consider the interaction of an incoming particle (atom or
nucleus having the ground state  $E_{g}$ and the excited state
$E_{e}$) with the crystal target. It is possible to choose the
conditions \cite{OKO65,OKO65A} when the frequency of a collision
particle with the atoms of crystal $\nu_{col}=V_{0}/a_{0}$ (the
velocity $V_{0}$ of a particle motion $a_{0}$ is the distance
between the atoms in the crystal) will be commensurable with the
transition frequency $\nu_{tr}$ of the particle
$$\nu_{tr}=\frac{E_{e}-E_{g}}{h}=\frac{n_{1}}{n_{2}}\nu_{col},$$
where $n_{i}$ - integer numbers. It is clear that at such
conditions the interaction between the particle and atoms of the
crystal should have a resonance character.

If the particle interacts with the $n$ atoms of the crystal, then
the probability to excite the particle is equal to
$$W(n)=W(1)n^{2},$$
where $W(1)$ is the probability of excitation of the particle by
one atom of the crystal.

This is a collective (coherent) amplification mechanism of the
excitation for the projectiles in the periodic field of the
crystal predicted first by Okorokov V.V. \cite{OKO65,OKO65A} and
observed experimentally by Okorokov V.V. too
\cite{OKO72,OKO73,OKO02}.

The resonance and coherent amplification of atoms and nuclei
excitations by the periodic fields of crystals is now well
established and recognized by the physical society and is used in
different applications but is not known for the cold fusion
society.

>From a modern point of view water has a very complicated
geometrical structure as a collection of quasicrystal clusters (
see \cite{SEN04} and references in it). The hydrogen atom, atoms
and molecules, water and solutions, solid states and condensed
matter have the same homology in the geometric structure, where
the de Broglie wave length of electron in the ground state of
hydrogen atom plays the role of the standard one \cite{GAR00}.

The puzzle of poor reproducibility of experimental data of LENR is
now evident:
\begin{itemize}
\item Electrolysis in solutions, discharge in gases and any
external influence on atoms leads to\\
1.The atoms are ionized, thus changing radioactive rates by bound
state $\beta_{b}$-decay of nuclei.\\
2. The ions can be accelerated in a such way that they come to
resonance conditions to intensify  excitation of
nuclei, atoms,...\\
3. Even small external fields can induce large responses as an
avalanche in the mountains is stimulated, say, by an accident cry.
\end{itemize}

 A mechanical analogue of the observed phenomenon is
the synchronization of oscillations of the pendulum clock
suspended from the moving girder -- the Huygens synchronization
principle \cite{BLE81}. The universal resonance synchronization
principle for a microsystem (for nuclei, atoms, molecules for
living and nonliving sells,...) was established in \cite{GAR01}.

The decrease and increase radioactivity of tritium with increasing
temperature in small titanium particles was observed \cite{REI94}
whereas current experiment and theory overlooked this effect.

\subsection{Nuclei and Atoms as Resonators}

In 1953 Schwartz  H.M. in 1953 \cite{SCHWA} proposed the nuclear
and the corresponding atomic transitions be considered as a
unified whole process. This process contains the $\beta$ decay
which represents the transition of nucleon from one state to an
other with emission of electron and antineutrino and,
simultaneously the transition of atomic shell from the initial
state to the final one. A complete and strict solution of this
problem is still waiting for its time (see, for example, a review
paper \cite{KAR02}).

The division of decay energy into nuclear and atomic energies has
only a conditional sense, especially, in the resonance case. The
process has a resonance charter and its probability is large when
energy differences of nuclear and atomic transitions become close
to zero. The drastic acceleration of decay time in  $H$-like ions
of $^{229m}Th$ may be up to $10^{5}$ \cite{KAR02}, the electron
shell serves as a trigger, reducing the lifetime of the isomer by
up to five orders of magnitude. The fantastic acceleration of
decay time for the case $E3$-transition in $^{235}U$ may be up to
20 orders of magnitude.

The probability of the resonance transfer of energy by electrons
from the nuclei can be increased by application of laser, which
compensates the defect of resonance. The corresponding enhancement
factor in some cases  may be $10^{3}$. It is important to note
that the knowledge of isomer energy is not necessary, the laser
should be synchronized on the atomic frequency.

\begin{itemize}
\item This is a real phenomenon of resonance synchronization (see
Appendix) of nuclear, atomic, and laser frequencies  to control
the decay processes.
\end{itemize}

It is also predicted \cite{GAN84} that the lifetime of the
hindered photo-fission can be reduced up to $10^{3}-10^{4}$ by
application of laser. Laser in a such case changes the angular
momentum of a decaying state by unity practically without altering
its energy.

\begin{itemize}
\item Low-energy external fields in LENR can play a role of a
trigger changing the quantum numbers of the hindered or forbidden
processes so that the first should  be enhanced  and the second
should be allowed. This mechanism inducing LENR may be the main
reason for  poor reproducibility of LENR experiments and main
mechanism of geo- and biotransmutations.
\end{itemize}

\subsection{Geo-, Bio- and Alchimical Transmutations}

All the above-described mechanisms of LENR are grounded on the
universal resonance synchronization principle (see Appendix). The
main requirement of this principle is that the frequencies
$\nu_{i}(ext)$ of external fields should be
$\underline{commensurable}$ with the frequencies $\nu_{j}(in)$ of
subsystems making a whole system:
$\nu_{i}(ext)=\nu_{j}(in)n(j)/n(i).$ We strongly emphasize that
the frequencies of an external field can be
$\underline{infinitly}$ $\underline{small}$  in comparison with
the corresponding frequencies of subsystems. The frequencies
$\nu_{i}(ext)$ are as  triggers starting  emission of internal
energy. The enhancement (resonance) effect on LENR induced by
external fields can be extremely large (small) when  maximal
values of density distributions for external fields and the
corresponding distributions of a subsystem coincide (do not
coincide). It means that:

\begin{itemize}
\item Even extremely low-energy external fields may induced
nuclear transmutations with emission of internal high energies,
according to the universal resonance synchronization principle.
\end{itemize}

Natural geo-transmutations in the atmosphere and earth occur at
the points  of strong change in geo- and electromagnetic fields
\cite{KRI03,JON03,MAM81}. V.I. Vysotsky and A.A. Kornilova
published an excellent book: «Nuclear Fusion and Transmutation of
Isotopes in Biological System» \cite{VYS03}, we refer a reader to
this book.

It seems (for F.A.Gareev) that the fact that some alchemists may
\cite{JAO02} «change base metals into nobel ones, silver or gold»
does not contradict  the mechanisms of LENR described above.

\subsection{ Ball lightning as a macroscopic low energy nuclear
reactor}

       All internal contradictions of the previous theories of a ball lightning
       were based, by default, on an assumption that the ball lightning is a plasmoid.
       In order to maintain the macroscopic volume of air (the mixture of nitrogen,
       oxygen, water vapour, etc.) in ionized condition, it is necessary to provide a
       great amount of energy from some kind of a source. Many experimenters,
       among them are such well-known experts as P.L. Kapitsa, made repeated
       attempts to create a long-living spherical plasmoid in laboratory conditions.
       However, no efficient ways of supplying the isolated plasma clots with energy
       and maintaining them in a stationary condition for a few minutes (that is
       the lifetime of a natural ball lightning) could be found.

       The purpose of this paper is to substantiate a hypothesis that the natural
       ball lightning is an area of space where the chain nuclear reaction of the
       bound-state $\beta$-decay of radioactive phosphorus nuclei takes place. It is
       shown that the analyzed phenomenon is related to the physics of electrical
       discharge in gases indirectly. Therefore, the term globular lightning is not
       sufficiently correct.

       The main hypothesis which is asserted hereinafter was formulated for the
       first time in \cite{RAT}.  The logic of the creation of this hypothesis is as follows:
\begin{enumerate}
\item  Ball lightning always leaves a smell of sulphur, ozone, and
nitrogen oxide after itself \cite{SMI90}.
 \item  Sulphur can be generated only as a result of phosphorus $\beta$-decay \cite{SEL70}.
 \item Rate constant of $\beta$-decay depends on a lot of the ionization degree of decaying
radionuclide [33]. The half-life of ionized radiophosphorus is
approximately 1-2 minutes and is comparable with the lifetime of
ball lightning in natural conditions.
 \item Radiophosphorus is abundant in nature. It is found in rain-water in macroscopic
amounts \cite{LAL57}
\end{enumerate}

       This model was proved in \cite{RAT} and we can say that it is now confirmed.
       Thus, it is a type of a natural low-energy nuclear reactor.

\subsection{Demkov and Meyer super-focusing}

The ion flux transmission through a monocrystalline  medium is
accompanied by many interesting and unexplained phenomena (see  a
review papers \cite{GIB75,KAR94}). Among them we indicate the
effect of "channeling" in crystals: the enhancement and reduction
of flux near crystalloghraphic directions. Yu.N. Demkov and J.D.
Meyer \cite{DEM03} propose to use a "channeling" effect in the
stimulation and enhancement of LENR by following way:

 "A highly collimated beam of
protons ($\approx$  1 MeV) entering the channel of a monoctystal
film forms at a certain depth an extremely sharp (<0.005 nm)  and
relatively long  (some monolayers of the crystal) focusing area
where the increase of the flux can reach thousand times. Impinging
atoms in this focusing area can undergo nuclear reactions with
proper foreign dopants  which disappear if the crystal is tilted
from this position by only   radians. This effect can be called
super-focusing in the channels, in contrast to the ordinary flux
peaking where the increase of flux reaches only few times. Results
are confirmed by the Monte Carlo calculations accounting for
several properties of the real lattice."

\section{Conclusions}

We have concluded that LENR is possible in the framework of the
modern physical theory - the universal resonance synchronization
principle and based on it different enhancement mechanisms of
reaction rates are responsible for it\footnote{Intensification of
LENR using superwave excitation \cite{DAR03} is based on this
principle.}. Investigation of this phenomenon requires the
knowledge of different branches of science: nuclear and atomic
physics, chemistry and electrochemistry, condensed matter and
solid state physics,... The results of this research field can
provide a new source of energy, substances, and technologies.

The puzzle of poor reproducibility of experimental data is due to
the fact that LENR occurs in open systems and is extremely
sensitive to parameters of external fields. Poor reproducibility
and unexplained results do not mean that the experiment is
wrong\footnote{Solutions of salts, electrolytes and living systems
contain a large amount of ions. In these cases the bound state
$\beta_{b}$-decay and other described above enhancement mechanisms
of LENR can play an essential role. Unfortunately, we do not know
the works devoted to this problem.}.

\section{Appendix}
\begin{center}
\bf{The Universal Resonance Principle of Synchronization}
\end{center}

Many objects in Nature - elementary particles, nuclei, atoms,
molecules,..., DNA, proteins, etc. are built as self-consistent
hierarchical systems and have the same homological constructions
in the sense that they are found by the same fundamental physical
laws: energy-momentum conservation law and sectorial conservation
law (the second Kepler law). Schrodinger \cite{1} wrote that an
interaction between microscopic physical objects is controlled by
specific resonance laws. According to these laws, any interaction
in a microscopic hierarchic wave system exhibits the resonance
character. The difference between eigenenergies (eigenfrequences)
in one system should be equal to each other
\begin{equation}
h \nu_1 - h \nu_{1}^{'} = h \nu_{2}^{'} - h \nu_{2},\;\;\; \nu_1 - \nu_{1}^{'} = \nu_{2}^{'} - \nu_{2}. \label{eq:eq1}
\end{equation}
Therefore, eigenfrequencies are additive. In other words, the
resonance condition is formulated in the following way:
oscillations participating in an interaction process should be
constituents of the same frequencies. Thus, we come to the
important conclusion: in the whole interacting self--consistent
wave system the hierarchy of frequencies is established. So the
sum of all partial frequencies is the integral of motion. Due to
the above-said, the corresponding partial motions are determinate.
This determinism arises as a consequence of  the energy
conservation law. As the resonance condition arises from the
fundamental energy conservation law, the rhythms and
synchronization of the majority of phenomena to be observed are
the reflection of the universal property of self--organization of
the Universe. The resonance synchronization principle is
substantiated at the microscopic level (see, for details \cite{2})
as the consequence of energy conservation law and resonance
character of any interaction between wave systems. In this paper,
we have demonstrated the universality of the resonance
synchronization principle independent of substance, fields and
interactions for microsystems. Thereby, we bring some arguments in
favor of the mechanism - ORDER from ORDER, declared by Schrodinger
in \cite{3}, fundamental problem of contemporary science. We come
to a conclusion \cite{4} that a stable proton and  a neutron play
the role of a standard for other elementary particles and nuclei.
They contain all necessary information about the structure of
other particles and nuclei. This information is used and
reproduced by simple rational relations, according to the
fundamental conservation law of energy-momentum. We originated the
principles of commensurability and self-similarity \cite{5}. The
commensurability and self--similarity result in the very unity of
the world.  The principle of commensurability is displayed in
phenomena in different branches of science \cite{5}.

All material objects (micro-- and macrosystems), which are
described by standing waves, know all about each other. Each
object is the scaled one of the other and it is not possible to
say which is more «fundamental». In this work, we have
demonstrated that the structure of DNA and cell molecules can be
calculated with some  structure  of a hydrogen atom. The
interatomic distances in cell molecules are quantized according to
the quantization rule of the fractional Hall effect. Therefore, we
can conclude that the structure of DNA and cell molecules can be
established from the analysis of hydrogen spectra using the
quantization rule of the Hall effect and vice versa \cite{6}. The
bridge between the structure of a hydrogen atom, cell molecules
and the Hall effect exists! It is very surprising that there are
phenomena in Nature that are really described by simple rational
relations. Only the fundamental conservation law of energy --
momentum is responsible for this harmonic movement.

The resonance principle of synchronization became a fruitful
interdisciplinary science of general laws of self-organized
processes in different branches of physics. It is intriguing to
speculate that  many questions can now be  formulated as a result
of universality of the resonance synchronization principle
independent of substance, fields, matter and interactions for
micro- and macrosystems \cite{6}. Information concerning important
details of an ecosystem's evolution is contained in frequency
spectra. Therefore, matter turns out to be a form of organized
information. The Universe was arranged according to number,
harmony, and perfect forms.

A new concept in evolution is robustness. One suggests simulating
evolution of complex organisms constrained by the sole requirement
of robustness in their expression patterns. Robustness in
biophysics is defined as the ability to function in face of
substantial changes in components. Robustness is implemented by
constraining subsequent patterns to have similar expression
patterns. Key properties of biochemical networks are robust, i.e.,
they are insensitive to  precise values of the biochemical
parameters \cite{7}. Robustness is an important ingredient in
simple molecular networks and, probably, also an important feature
of gene regulation.  S. Bornhold and  K.  Sneppen \cite{8} suggest
considering robustness as an evolutionary principle. We came to
the conclusion that the robustness principle can be understood in
the framework of the universal resonance synchronization
principle.

We have concluded that the homology of atom and molecule
structures exists. It means that the de Broglie wave length
$\lambda_{e}$ of electron in the ground state of a hydrogen atom
plays the standard role -- all interatomic distances in molecules
could be commensurable with  $\lambda_{e}$. There are huge
examples of commensurable ratios between the interatomic distances
and  $\lambda_{e}$ in superconducting, nanomaterials, DNA,
protein, $\dots$, living molecules \cite{6}.

A molecule is an aggregate of atoms in a distinct three
dimensional arrangement. Distances between atoms fix the structure
of the molecules, as was so forcefully emphasized by L. Pauling.
These interatomic distances depend on the resonance interactions
between atoms and also on the  sizes of atoms. We come to the
conclusion that each object in the hierarchical system is scaled
one of the other and it is impossible to say which is more
«fundamental». We assume now, as a working hypothesis, that the De
Broglie electron wave length in a hydrogen atom in the ground
state can be considered as a standard  of dimensions for atoms and
interatomic distances in molecules. So interatomic distances and
radii of atoms can be written in the following way:
\begin{equation}
R=\frac{n_{1}}{n_{2}}\lambda_{e}, \label{eq:eq2}
\end{equation}
where $\lambda_{e} = 0.33249185$ nm is the de Broglie electron
wave length in a hydrogen atom in the ground state and
$n_{1}(n_{2})=1,2,3,...$.

\begin{itemize}
\item Note that the quantization conditions for the fractional Hall
effect \cite{6} are the same as (\ref{eq:eq2}). It means that the
fractional Hall effect demonstrates only the commensurable
velocities of electrons in hydrogen atoms and $GaAs$-type
heterostructures (two-dimensional electron gas). So there  is no
room for interpretation of the fractional  Hall effect in terms of
the fractional charge. Nobody observed the fractional charge in
Nature.
\end{itemize}

It is well known in optics (in quantum mechanics too) that the
transition coefficient of light through the layer is equal to one
if the following relations between the thickness $R$ of the layer
and wave length $\lambda_e$  exist
\begin{equation}
R = \frac{n}{4} \lambda_e, \;\;\;n = 1, 2, 3, \dots \;.    \label{eq:eq3}
\end{equation}
It is interesting to note that: $1)$ the Bohr quantization
conditions $\lambda_N = N \lambda_e$ for a hydrogen atom and the
quantization conditions $\lambda_N = N \lambda_{^4 He}$ for
superfluid $^4 He$ coincide with (\ref{eq:eq3}) if $N = n / 4$;
$2)$ the Tomasch quantization conditions for tunneling are the
same as (\ref{eq:eq3}).

We have carried out a systematic analysis of interatomic distances
for a huge number of systems, using (\ref{eq:eq3}), in which
$\lambda = \lambda_e$ is the electron wave length in the ground
state of a hydrogen atom. We came to the conclusion that the
superconductivity can be explained by the assumption: channel
motions in  systems like that and electron motion in the ground
state of a hydrogen atom are exactly synchronous. Therefore,
superconductivity systems represent a coherent synchronized state
-- complex of coupled resonators with the commensurable
frequencies.

\begin{itemize}
\item  It means that we have in principle found out the possibility to achieve
superconductivity at room temperature \cite{6}.
\end{itemize}

The parameter--free formula for interatomic distances in
biomoleculas, superconducters, and size of nanostructures has been
obtained. This establishes some bridge between the structures of
different phenomena (conductivity, superconductivity,
insulator--metal transmission, quantum Hall effect, superfluidity,
quantization of nanostructure cluster size, size of biomolecules).
This connection can be considered as an indication of existence of
some physical phenomena in the structures of the superconducting
and living systems.

We have shown \cite{6} only a small part of our calculations by
formula (\ref{eq:eq3}) and the corresponding comparison with
experimental data for interatomic distances in some molecules. One
can be surprised by a high accuracy description of the existing
experimental data.

Understanding of the origin and evolution of the genetic code must
be the basis for a detailed knowledge of the relationship between
the basic building blocks of DNA and environment. As is widely
accepted today, essentially all the DNA in an eukaryotic nucleus
are formed  of histones and different chromatin structures folded
hierarchically. At least five orders of DNA and chromatin
organization and folding (nucleotide, helix, nucleosome, solenoid
and chromatin fibre loop) have been described in literature. A DNA
chain is a long unbranched polimer composed of only four types of
subunits. These are nucleotides containing the basis adenin (A),
cytozine (C), guanin (G), and thymine (T). These nucleotides form
complementary flat pairs and the distances between these plains
are equal to $\lambda_e$.

\begin{itemize}
\item It means that the structures, formed in DNA molecules by
nucleotides, produce the two-- and three--dimensional waveguide.
\end{itemize}

All proteins look like dimers in which the two copies of the recognition
helix are separated by exactly one turn of the DNA helix: $3.4$ nm $\Leftrightarrow 10  \lambda_e = 3.325$ nm.

The DNA is packaged with histones into regularly repeating
nucleosomes that are packed into 30 nm (it's diameter) fibers;
$30 = 90 \lambda_e = 29.92$ nm, it is also elaborated folded and
organized by other proteins into a series of subdomains of distinct
character. This higher--order packing is the most fascinating and also
most poorly understood aspect of chromatin.

Molecules of DNA, amino acids, proteins, $\dots$ contain
tetrahedral blocks $H_3 C - C$ with the angles $<HCH =\; <HCH =
109.47^{\circ}$, with the bond length $3 d (H - C) = \lambda_e =
0.3325$ nm and $ 3 d (H - C) + d (C - C) = 3/2 \lambda_e = 0.4
987$ nm. Note that these molecules of amino acids and DNA have
planar blocks $H_2 N - C$, whose bond length is equal to $2 d (H -
N) + d (N - C) = \lambda_e = 0.332$ nm. Pentagonal rings in adenin
and guanin  have the bond length equal to $0.668$ and $0.666$ nm,
respectively, which is close to $2 \lambda_e = 0.665$ nm.

Many distances in living molecules are commensurable with the de
Broglie wave length $\lambda_e$ of an electron in the ground state
of a hydrogen atom. This means that $\lambda_e$  play the role of
the standard distance in molecules, especially in living
molecules. Hence, the electron motions in a hydrogen atom and in
living molecules are synchronized and self-consistent. A hydrogen
atom represents radiating and accepting  antennas swapping the
information with the surrounding substance.

M. Gryzinski  \cite{greez,GRY04,GRY75} has proved that atoms are
the quasi--crystal structure with  definite angles: $90^\circ,
109^\circ \; and \; 120^\circ$, which are the  well-known angles
in crystallography.

\begin{itemize}
\item We have proved the homology of atom, molecule, and crystal structures.
\end{itemize}

So the resonance synchronization principle  is substantiated at the microscopic  level
as the consequence of the energy conservation law and resonance character of any
interaction between wave systems. The commensurability and self-similarity result
in the very unity of the world.

It means that our method can be used in  different fields of
fundamental researches and also in applications: construction of
new materials, say, high--temperature superconductors, new drugs
in medicine, new methods in diagnostics of diseases, and new
devices by analogy with biomolecules.

\begin{center}
\bf{Atoms as open systems}
\end{center}

The conservation laws fulfill for a closed systems. Therefore, the
failure of parity in week interactions means that the
corresponding systems are the open systems. Periodic variations
(24 hours, 27, and 365 days \cite{SHN98,BAU95,BAU98,BAU01}) in
$\beta$-decay rates indicate that failure of parity in week
interactions have a cosmophysical origin. The charged particles
moving with acceleration should radiate (absorb) electromagnetic
waves - the fundamental classical electrodynamics low. The stable
orbits of electrons in atoms are exist,  but electrons do not
radiate on them according to third Bohr's postulate (third Bohr's
postulate in 1913 - "Despite the fact that it is constantly
accelerating, an electron moving in such an allowed orbit does not
radiate electromagnetic energy. Thus, its total energy E remains
constant.").  Why electrons do not radiate on the stable state of
atoms - nobody knows it. We formulate as a working hypothesis the
following assumptions:
\begin{itemize}
\item The classical lows of physics are valid for macro- and
microsystems. Contradiction between classical electrodynamics and
quantum theory should be solved a very simple way.  Proton and
electron in hydrogen atom move with the same frequency, their
motions are synchronized. A hydrogen atom represents radiation and
accepting antennas (dipole) interchanging of energy with the
surrounding substance. This energy is the relict radiation energy.
\item The relict radiation (T=2.725 K) should play a role of
conductor for proton and electron motions in the hydrogen atom due
to the universal resonance synchronization principle. The external
field - relict radiation field and hydrogen atom form an
auto-oscillation system in which the frequencies of an external
field and frequencies of the whole subsystem are commensurable.
The demand for frequency commensurability means that all motions
are in a co-ordination (in resonance).
\item The sum of radiate
and absorb energies by electron and proton moving in an allowed
orbit is equal to zero. THUS, ITS TOTAL ENERGY E REMAINS CONSTANT
- only the last part of the third Bohr's postulate is correct.
\item The relict radiation is a result of  the selforganization of
stable hydrogen atom according    to the universal resonance
synchronization principle.
\end{itemize}

\end{document}